\def\be{\begin{equation}}
\def\ee{\end{equation}}
\def\bea{\begin{eqnarray}}
\def\eea{\end{eqnarray}}
\documentclass[12pt]{iopart}
\usepackage{epsfig,graphicx}

\begin{document}

\title[Search for Squeezed-Pair Correlations at RHIC]{Search for Squeezed-Pair Correlations at RHIC}

\author{Sandra ~S. ~Padula$^1$,  O.~Socolowski Jr.$^2$,  T.~Cs\"org\H{o}$^3$, and M.~I.~Nagy$^3$ }

\address{$^1$Instituto de F\'\i sica Te\'orica-UNESP, 01405-900 S\~ao Paulo, SP, Brazil\\
$^2$Campus Experimental de Itapeva, UNESP, 18409-010 Itapeva, SP, Brazil\\
$^3$MTA KFKI RMKI, Budapest, H - 1525, Hungary} 

\begin{abstract}
Squeezed correlations of particle-antiparticle pairs, also called Back-to-Back Correlations, are 
predicted to appear if the hadron masses are modified in the hot and dense hadronic 
medium formed in high energy nucleus-nucleus collisions. Although well-established theoretically, the 
squeezed-particle correlations have not yet been searched for experimentally in high energy hadronic 
or heavy ion collisions, clearly requiring optimized  forms to  experimentally search for this effect.  
Within a non-relativistic treatment developed earlier we show that one promising way to 
 search for the BBC signal is to look into the squeezed correlation function of pairs of $\phi$'s 
 at RHIC energies, plotted in terms of the average momentum of the pair,  $\bf {K_{12}}\!\!=\!\!\frac{1}{2} {\bf (k_1 + k_2)} $.  This variable's modulus, $2 |{\bf K_{12}}|$, is the non-relativistic limit of the variable $Q_{bbc}$, introduced herewith.  The squeezing effects on the HBT correlation function are also discussed.

\end{abstract}


\section{Introduction}

In the late 90's, Back-to-Back Correlations (BBC) among boson-antiboson pairs were 
established as theoretically expected signatures of 
in-medium mass modification of hadrons freezing out from a hot and dense medium\cite{acg99}. 
Shortly after, similar BBC was predicted to exist among fermion-antifermion pairs with in-medium modified masses\cite{pchkp01}. The fermionic (fBBC) and the bosonic (bBBC)  Back-to-Back Correlations can be treated by analogous formalisms, being both positive correlations with unlimited intensity. Their similar behavior is in contrast to what is observed in 
quantum statistical correlations of identical hadrons  (the HBT effect), where bosons with similar momenta have positive correlations, while fermions with similar momenta are anti-correlated.

The two-particle correlation function is defined as 
$ C_2({\mathbf k}_1,{\mathbf k}_2)\!=\!\frac{N_2(\mathbf k_1,\mathbf k_2)} {N_1(\mathbf k_1) \,
N_1(\mathbf k_2) }, \label{defC2} $
where $N_2(\mathbf k_1,\mathbf k_2)\!=\!\omega_{\mathbf k_1} \omega_{\mathbf k_2} \, \langle
a^\dagger_{\mathbf k_1} a^\dagger_{\mathbf k_2} a_{\mathbf k_2} a_{\mathbf k_1} \rangle$ 
 and 
 $N_1(\mathbf k_1)\!=\!\omega_{\mathbf k_1} \frac{d^3N}{d\mathbf k_1} \!=\!
\omega_{\mathbf k_1}\,
\langle a^\dagger_{\mathbf k_1} a_{\mathbf k_1} \rangle $, 
being $a^\dagger_\mathbf k$ and $a_\mathbf k$ the free-particle creation and
annihilation operators of scalar quanta, and 
$\langle ... \rangle $ means thermal averages. 
A generalization
of Wick's theorem for {\em locally} equilibrated 
systems in \cite{gykw} allows the factorization $N_2(\mathbf k_1,\mathbf k_2)\!=\!\omega_{\mathbf k_1} \omega_{\mathbf k_2} \, \Bigl[\langle
a^\dagger_{\mathbf k_1} a_{\mathbf k_1}\rangle \langle a^\dagger_{\mathbf k_2}
a_{\mathbf k_2} \rangle \pm \langle a^\dagger_{\mathbf k_1} a_{\mathbf k_2}\rangle
\langle a^\dagger_{\mathbf k_2} a_{\mathbf k_1} \rangle + \langle
a^\dagger_{\mathbf k_1} a^\dagger_{\mathbf k_2} \rangle \langle a_{\mathbf k_2}
a_{\mathbf k_1} \rangle\Bigr]$, where 
the {\small $(+)$} sign refers to boson as the {\small $(-)$} sign to fermions. 
The annihilation (creation) operator of the asymptotic, observed bosons with momentum 
$k^\mu\!=\!(\omega_k,{\bf k})$, $a$ ($a^\dagger$),  is related to the in-medium annihilation (creation) 
operator $b$ ($b^\dagger$), corresponding to thermalized quasi-particles, by the Bogoliubov-Valatin 
transformation,  
$a_k=c_k b_k + s^*_{-k} b^\dagger_{-k} $ ($a^\dagger_k=c^*_k
b^\dagger_k + s_{-k} b_{-k}$),
where $c_k=\cosh(f_k)$ and $s_k=\sinh(f_k)$; 
$ f_{i,j}(x)=\frac{1}{2}\log\left[\frac{K^{\mu}_{i,j}(x)\, u_\mu
(x)} {K^{*\nu}_{i,j}(x) \, u_\nu(x)}\right] $ is called the {\sl squeezing
parameter}, since the Bogoliubov-Valatin transformation creates squeezed
states from coherent ones. 
Introducing the chaotic and squeezed amplitudes, respectively, as
$ G_c({\mathbf k_1},{\mathbf k_2}) = \sqrt{\omega_{\mathbf k_1} \omega_{\mathbf k_2}} \; \langle
a^\dagger_{\mathbf k_1} a_{\mathbf k_2} \rangle,  
G_s({\mathbf k_1},{\mathbf k_2}) = \sqrt{\omega_{\mathbf k_1} \omega_{\mathbf k_2} } \; \langle
a_{\mathbf k_1} a_{\mathbf k_2} \rangle$, 
the bosonic two-particle correlation function can be written as
\be C_2({\mathbf k_1},{\mathbf k_2}) =1 + 
\frac{|G_c({\mathbf k_1},{\mathbf k_2})|^2}{G_c({\mathbf k_1},{\mathbf k_1})
G_c({\mathbf k_2},{\mathbf k_2})} + \frac{|G_s({\mathbf k_1},{\mathbf k_2})
|^2}{G_c({\mathbf k_1},{\mathbf k_1}) G_c({\mathbf k_2},{\mathbf k_2}) }, \label{fullcorr}
\ee where the first two terms correspond to the HBT correlation, whereas the first and the last terms represent the correlation function between the particle and its antiparticle, i.e., the squeezing part. In cases where the particle is its own anti-particle (for
$\pi^0$$\pi^0$ or $\phi$$\phi$ boson pairs, for instance), the 
full correlation function is to be considered, as in (\ref{fullcorr}).
The in-medium modified mass, $m_*$, is related to the 
asymptotic mass, $m$, by
$m_*^2(|{\bf k}|)=m^2-\delta M^2(|{\bf k}|)$, here assumed to be a constant mass-shift. 
For a hydrodynamical ensemble, both the chaotic and the squeezed
amplitudes, $G_c$ and $G_s$, respectively, can be written in a 
special form derived 
in \cite{sm1,sm2}, and developed in \cite{acg99,phkpc05}. 

\section{\bf Recent results} 
 
The formulation was initially derived for a static, infinite medium \cite{acg99,pchkp01}. More  recently, it was shown\cite{phkpc05}  that, in the case of finite-size systems expanding with moderate flow,  the squeezed correlations may survive with sizable  strength to be observed experimentally. A non-relativistic treatment with flow-independent squeezing parameter was adopted for the sake of simplicity, which allowed to obtain analytic results. The detailed discussion is in Ref. \cite{phkpc05}, where the maximum value of $C_s({\mathbf k},-{\mathbf k})$, was studied as a 
function of the modified mass, $m_*$, for strict back-to-back pair-momentum, 
${\mathbf k_1}\!\!=\!\!-{\mathbf k_2}\!\!=\!\!{\mathbf k}$, analogously to focusing on 
the intercept of the HBT correlation function. In practical terms, however, this was not very helpful to look for the BBC's experimentally,  
as two measured momenta are never exactly 
back-to-back. Therefore, it was necessary to combine the particles' momenta to clearly define variables in terms of which the BBC should be searched for. 

Within the non-relativistic approach of  \cite{phkpc05}, one possibility would be to  
select the particle-antiparticle pair momenta $(\mathbf k_1,\mathbf k_2)$, 
combine them as $\mathbf K\!=\!\frac{1}{2}( \mathbf k_1+\mathbf k_2)$, and analyze the squeezed correlation function in terms of $|\mathbf K|$, similarly to what is done in HBT. 
Since the maximum of the BBC effect is reached when ${\mathbf k_1}\!=\!-\!{\mathbf k_2}\!=\!{\mathbf k}$, this value would correspond to the limit $|\mathbf K|\!=\!0$. Therefore, the squeezed correlation should be investigated as 
$C_s({\mathbf k_1},{\mathbf k_2})=C_s({\mathbf 2 K},{\mathbf q})$. For simplicity,  in what follows we analyze the behavior of the correlation functions detailed in  \cite{phkpc05} by just taking values of $|\mathbf K|$ and $|\mathbf q|$.

Naturally, the analysis in terms of the variable $2\mathbf K$ would not be suited for a relativistic treatment. In this case, a relativistic momentum variable can be constructed, as $Q_{back}=(\omega_1-\omega_2,\mathbf k_1 + \mathbf k_2)=(q^0,2\mathbf K)$. In fact, it is preferable to redefine this variable  as $Q^2_{bbc} = -(Q_{back})^2=4(\omega_1\omega_2-K^\mu K_\mu )$, whose  non-relativistic limit is $Q^2_{bbc} \rightarrow (2 \mathbf K)^2$. 

\begin{figure}[htb]
\begin{center}
  \includegraphics[height=.24\textheight]{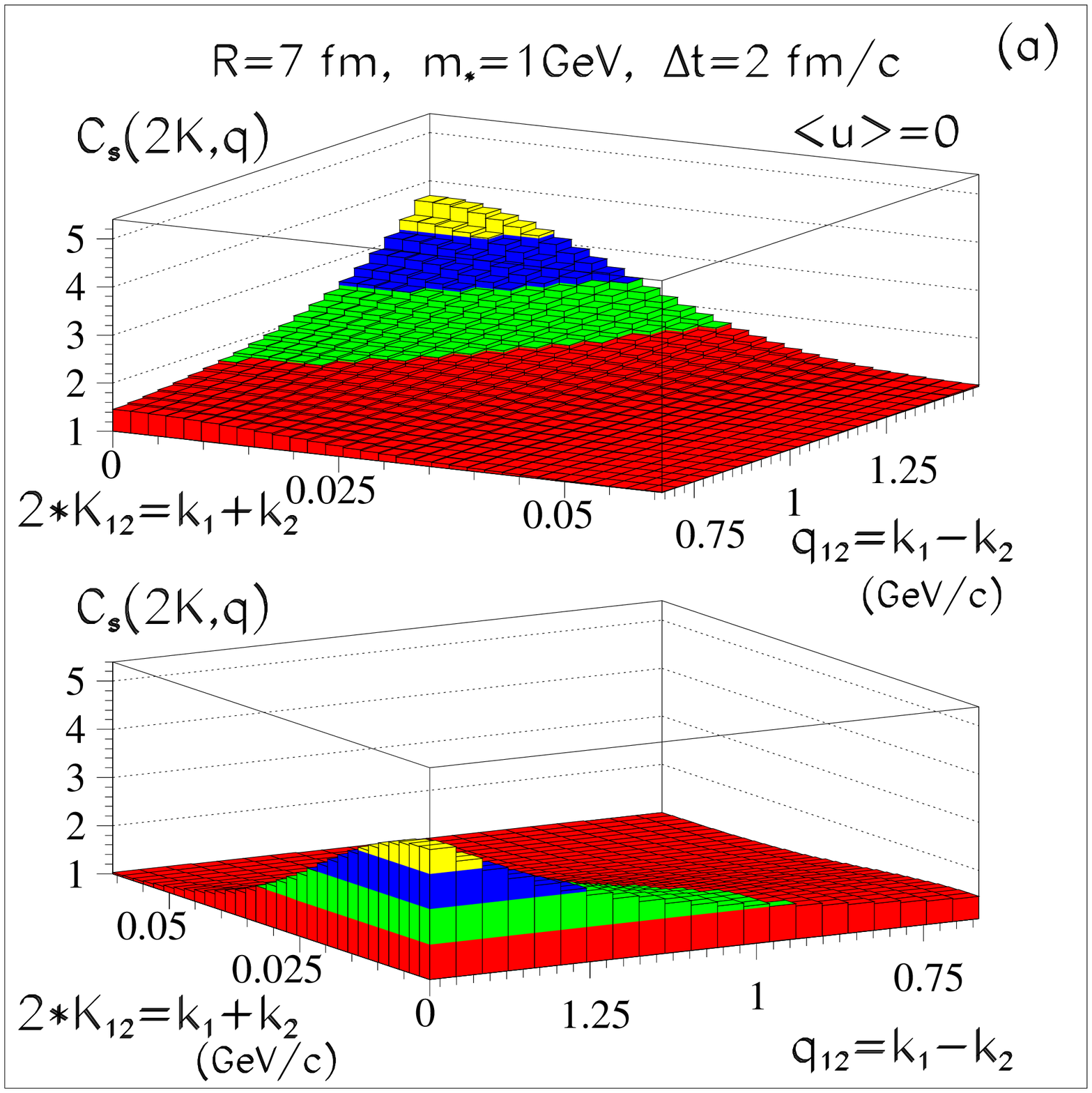}
  \includegraphics[height=.24\textheight]{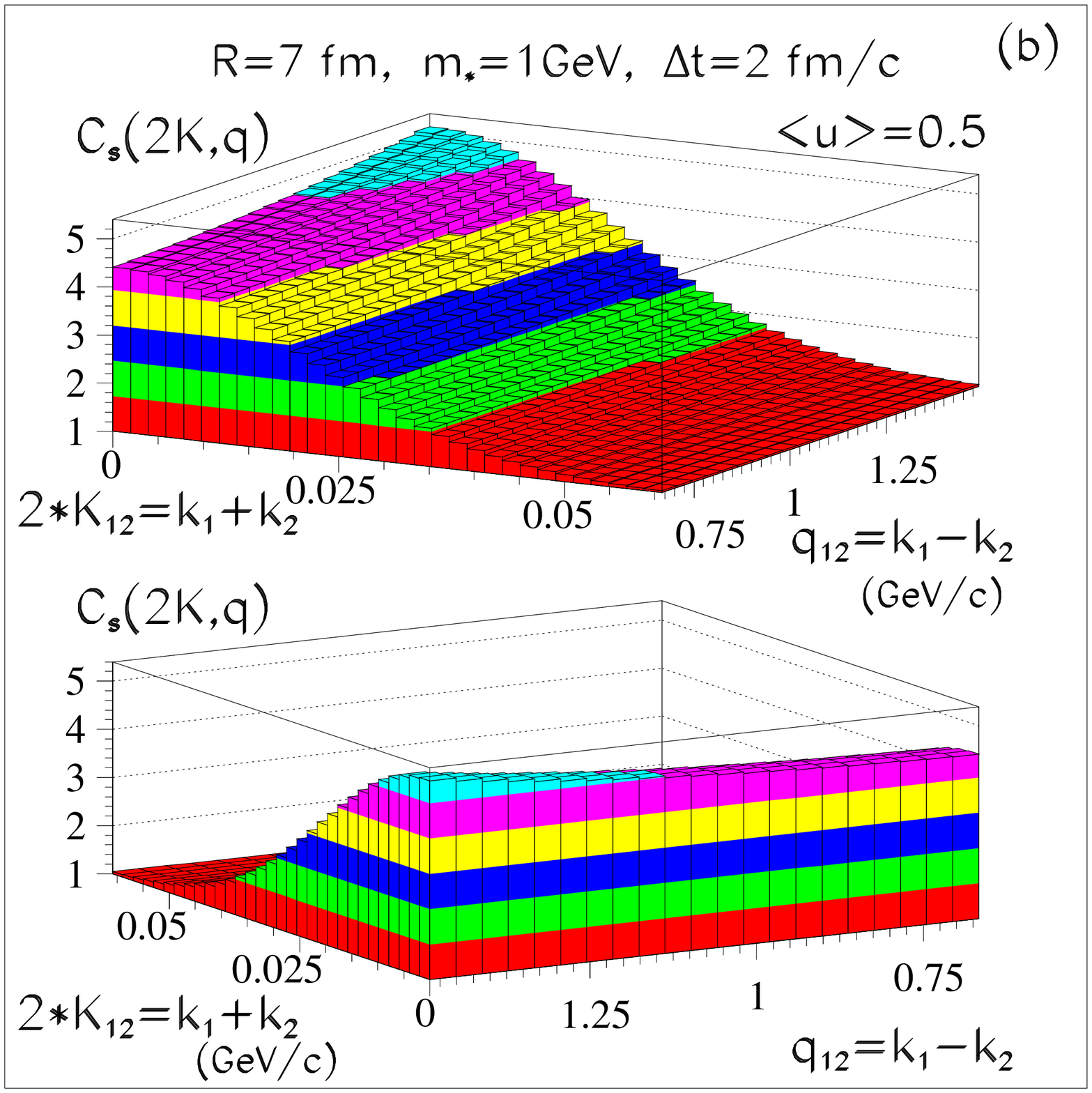}
 \caption{Squeezed-pair correlations are shown, illustrating the effects of flow, finite system size and finite emission time. In part (a) flow is absent, whereas (b) includes radial flow 
 ($\langle u \rangle\!=\!0.5$), which  enhances the BBC signal all over the investigated region.}
\end{center}
\end{figure}

%
By comparing the results in figure 1a and 1b we see that, in the absence of flow, the squeezed correlation signal grows faster for higher values $|\mathbf q|$ than the corresponding case in the presence of flow. However, this last one is stronger even for smaller values of  $|\mathbf q|$, showing that the presence of flow could help to enhance the signal. 

The sensitivity of the squeezed-pair correlation to the system size is shown in figure 2a for radii $R=7$ fm and $R=3$ fm, and is reflected in the inverse width of the curve as function of $2 |\mathbf K|$. 
In the absence of squeezing, i.e., if there is no in-medium mass modification, the squeezed correlation functions would be unity for all values of $2 |\mathbf K|$. 

\begin{figure}[htb]
\begin{center}
 \includegraphics[height=.25\textheight]{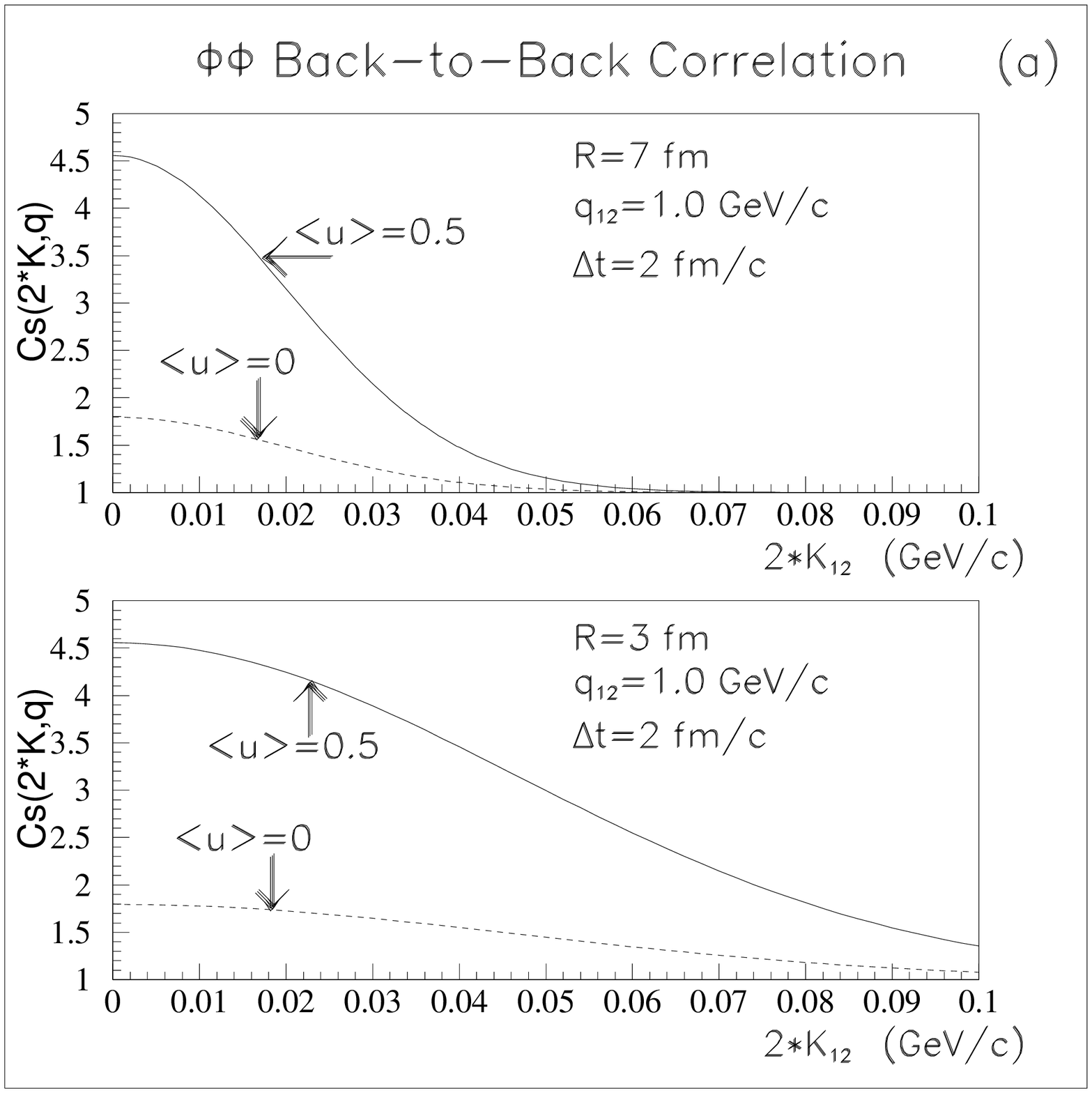}
 \includegraphics[height=.25\textheight]{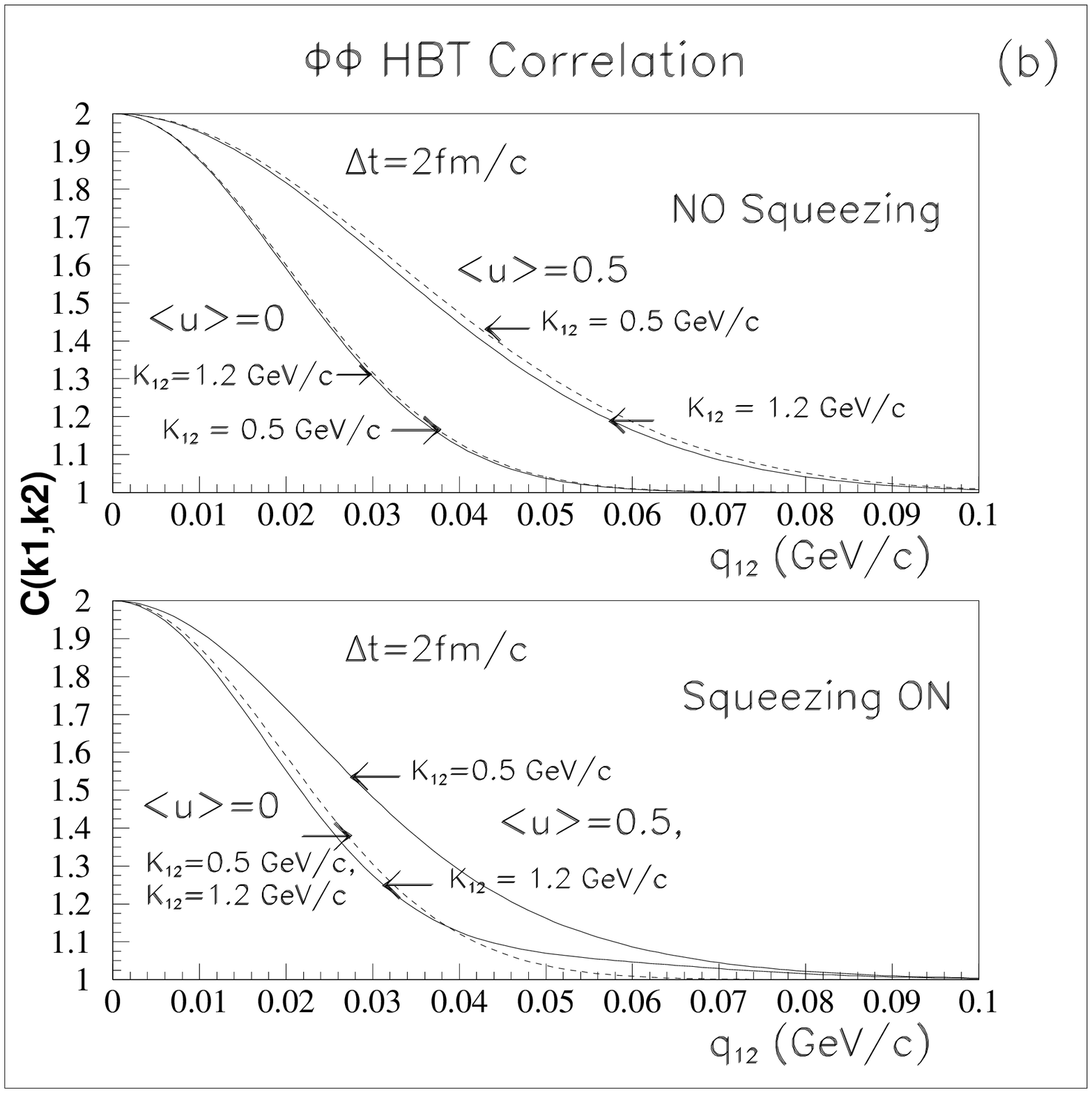}
\caption{Part (a) shows the response of the BBC function to system sizes with $R=7$fm (top) and $R=3$fm (bottom). Part (b) shows the HBT correlation function in the absence of squeezing (top part) and when it is present (bottom part). }
\end{center}
\end{figure}

The next question to be answered is how the behavior of the $\phi \phi$ HBT would be affected by the in-medium mass modification, since the HBT correlation function also depends on the squeezing factor, $f_{i,j}(m,m_*)$. 
For isolating the HBT effect from the BBC, we consider here the zone where the particle-antiparticle correlation is not significant and the HBT is the relevant contribution to (\ref{fullcorr}), as shown in figure 2b. 
Its top part shows only the effect of radial flow on the HBT correlation function, while in its lower part, both the flow and the squeezing effects are shown. We see that, without squeezing, the flow broadens the curves, as expected, whereas the squeezing effects 
tend to oppose to those of flow (for large  $|\mathbf K|$, it practically cancels the broadening of the correlation function due to flow), another striking indication of mass-modification, even in HBT!
\section{Conclusions}

In the present work we suggest an effective way to search for the Back-to-Back squeezed Correlations in heavy ion collisions at RHIC and later at LHC energies.  
Within a relativistic approach, we showed that a suitable variable to experimentally search for the squeezed correlation function is the average momentum of the pair, $(2 \mathbf K)^2$, which is the non-relativistic limit of the  proposed variable, $Q^2_{bbc}=4(\omega_1\omega_2-K^\mu K_\mu )$. 
We showed that, in the presence of flow, the signal is stronger over the momentum regions in the plots, i.e., roughly for $0 \le |\mathbf K| \le 100$ MeV and $500 \le |\mathbf q| \le 1200$ MeV, 
suggesting that flow may enhance the observable BBC signal. 
Another important point that we find within this simplified model and in the non-relativistic limit considered here is that the squeezing would distort significantly the HBT correlation function as well, tending to oppose to the flow effects on those curves, practically neutralizing it for large values of $|\mathbf q|$. Finally, it is important to note that all the effects and signals studied theoretically here would exist only if the particles analyzed had their masses modified by interactions in the hot and dense medium. If that would not happen, then the squeezed correlation functions would be flat in {\bf 1} and the HBT correlation functions would behave as usual, either in the presence or absence of flow. However, if the particles' masses are indeed modified 
in-medium, the experimental discovery of squeezed particle-antiparticle correlation (and the distortions pointed out in the HBT correlations) would be an unequivocal signature of 
those effects on hadronic probes, prior to freeze-out.  

\subsection{Acknowledgments}
SSP is very grateful to FAPESP for the support to attend the conference and to the QM2008 Organizing Committee for waiving her conference fee. T. Cs. was supported by the MTA-INSA exchange program,  and by OTKA grants T09466 and NK 73143.

\end{document}